\documentclass[pra,12pt,tightenlines,showpacs,superscriptaddress]{revtex4}

\usepackage{bm}
\usepackage{graphicx}

\begin{document}

\newcommand{\tr}{\,\text{tr}\,}
\newcommand{\re}{\,\text{Re}\,}
\newcommand{\im}{\,\text{Im}\,}
\newcommand{\s}{\,\text{s}}
\newcommand{\sech}{\,\text{sech}}
\newcommand{\sgn}{\,\mbox{sgn}\,}
\newcommand{\mod}{\,\mbox{mod}\,}
\newcommand{\Fix}{\,\mbox{Fix}\,}
\newcommand{\saw}{\,\mbox{saw}\,}

\title{Periodic orbit quantization of a Hamiltonian map on the sphere}

\author{A.~J.~Scott}
\affiliation{Department of Mathematics,}
\author{G.~J.~Milburn}
\affiliation{Centre for Quantum Computer Technology, \\The University of Queensland, QLD
4072 Australia.}

\begin{abstract}
In a previous paper we introduced examples of Hamiltonian mappings with phase 
space structures resembling circle packings. It was shown that a vast number of 
periodic orbits can be found using special properties. We now use this information 
to explore the semiclassical quantization of one of these maps.
\end{abstract}

\maketitle

\section{Introduction}\label{sectionI}

The general theory of periodic orbit quantization was first developed many years 
ago by Gutzwiller \cite{gutzwiller,gutzwiller2} and has ever since played an 
important role in our understanding of chaotic systems in the semiclassical 
regime. By applying stationary phase approximations to the trace of the Green's 
operator, Gutzwiller was able to derive a semiclassical formula for the density 
of states as the sum of a smooth part and an oscillating sum over all periodic
orbits of the corresponding classical system. Gutzwiller's `trace formula', 
however, is only valid for isolated periodic orbits which, in general, restricts 
his results to fully chaotic systems where the dynamics is completely hyperbolic.

At the other end of the scale lie integrable systems. Armed with the knowledge of 
all constants of motion one may obtain the semiclassical eigenenergies using 
EBK torus quantization \cite{EBK}. Alternatively, Berry and Tabor \cite{berry} 
showed that EBK quantization could be recast as a `topological sum' over the 
periodic orbits and obtained an analogue of Gutzwiller's trace formula for 
integrable systems. The Berry-Tabor formula has also been extended into the 
near-integrable regime \cite{ozorio,tomsovic}. However most physical systems are 
of a mixed type, exhibiting both regular and chaotic behaviour. Some progress on 
the periodic orbit quantization of mixed systems has also been made 
\cite{sieber,schomerus,main3}.

The theory referred to above is for autonomous systems. The dynamics of periodically 
driven systems and quantum maps is best described by the so-called Floquet 
operator \cite{haake}. Upon deriving an approximation for the trace of this 
unitary operator in terms of the classical periodic orbits, one may obtain a 
semiclassical approximation for the spectral density of its eigenphases. 
This trace formula has been derived for a class of maps on the plane 
\cite{tabor,junker}, on the torus \cite{laksh3} and for an array of special 
mappings with mixed results. The trace formula for Arnold's cat map is exact 
\cite{keating,boasman}. However the more general sawtooth map gives poor results 
\cite{laksh3,laksh2,sano}. It was found necessary to include boundary contributions
in the trace formula of the baker's map \cite{eckhardt,saraceno,laksh,toscano,tanner},
while ghost contributions needed to be included in the case of the kicked top 
\cite{kus,kus2,schomerus} and the standard map \cite{scharf,sundaram,scharf2,saito}.
Poincar\'{e} maps have also been considered \cite{bogomolny}. 

The model for which we will apply Gutzwiller's method of periodic orbit 
quantization is quite dissimilar to the above mappings. The dynamics of our 
model is dominated by an infinite number of isolated stable periodic orbits of 
arbitrarily long period. The resonances associated with these periodic orbits 
are circular disks and fill in the phase space in a manner resembling circle 
packings. Three of these mappings were introduced in a previous paper 
\cite{scott}, each corresponding to a particular phase space geometry: planar, 
hyperbolic and spherical. We will consider only the spherical map in this paper. 
Its quantum analogue lives in a finite dimensional Hilbert space allowing us to 
circumvent convergence problems associated with semiclassical trace formulae 
\cite{main2}. In addition, our model does not suffer from the exponential 
proliferation of periodic orbits found in most chaotic systems. The task of 
finding all periodic orbits of a given period is much simplified with the help 
of symbolic labeling and the use of special symmetries. Our mapping is not 
chaotic in a strict sense. However the presence of infinitely many stable 
periodic orbits with arbitrarily long period generates highly irregular motion. 
This behaviour is located on the borderline between regular and chaotic motion, 
but in a manner very different from the generic case of KAM systems. The strong 
deviation is explained by the mapping's non-smooth construction. 

Our motivation for considering such a system is as follows. Tests of the 
applicability of periodic orbit quantization as a theory which adequately 
describes quantum systems in their semiclassical limit has, for the most part, 
been focussed on fully chaotic systems. Systems with regions of stability have 
been ignored due to complications in the theory which arise whenever 
non-isolated periodic orbits are present. For integrable systems, the presence 
of periodic tori surrounding each stable periodic orbit is accommodated in the 
theory of Berry and Tabor. However when such a system is perturbed into the 
near-integrable regime, these tori resonate and form island chains of high-order 
stable and unstable periodic orbits of equal period. If the perturbation is 
small, the newly created periodic orbits will not be sufficiently isolated, and 
hence, Gutzwiller's theory cannot be applied. In such cases, uniform 
approximations need to be made which bridge the gap between the two theories 
\cite{ozorio,tomsovic}. For chaotic systems, each stable periodic orbit is 
locally near-integrable, and in general, will always be accompanied by island 
chains of non-isolated periodic orbits, which again inhibit the application of 
Gutzwiller's theory. To explore the limits of periodic orbit quantization for 
the case of stable motion we need to first investigate `toy systems' of low 
generality like ours where the method is expected to work best. The circular 
resonances associated with each stable periodic orbit of our mapping rotate 
about each point of the orbit in a linear fashion, and hence, contain no 
high-order resonances. For almost all parameter values, the period of rotation 
will be an irrational multiple of $\pi$. Thus, each stable periodic orbit 
will be isolated by the radius of its circular resonance, which, for 
a careful choice of the parameter values, can be made large enough such that 
Gutzwiller's theory is applicable. The low generality of our model should not 
render our research as being futile or purely academic. Indeed, the 
investigation of other toy systems, such as Sinai's billiard, has led to much 
insight for the hyperbolic case \cite{bohigas}. 

Our paper is sectioned as follows. In Section \ref{sectionII} we review the classical map and 
construct its generating function. For a more detailed exposition of the classical 
map one should refer to \cite{scott}. In Section \ref{sectionIII} we introduce the quantum map 
and spin coherent states. Using these states as a basis, we then derive semiclassical matrix 
elements of the Floquet operator in Section \ref{sectionIV}, followed by semiclassical 
traces and eigenphases in Section \ref{sectionV}. The theory in these sections 
follows the work by Ku\'s et al. \cite{kus}. Numerical investigations of our 
semiclassical approximations are contained in sections \ref{sectionVI}  and \ref{sectionVII}. Finally, in 
Section \ref{sectionVIII} we discuss our results.

\section{Classical map}\label{sectionII}

The Hamiltonian under consideration is that of a kicked linear top
\begin{equation}
H({\bf J},t)=\omega J_3+\mu|J_1|\sum_{n=-\infty}^{\infty}\delta(t-n), \label{H}
\end{equation}
where $\mu,\omega\in[0,2\pi)$ are parameters and $({\bf J})_i=J_i=\epsilon_{ijk}x_jp_k$ $(i=1,2,3)$ are the three components of angular momentum for a particle confined to a sphere, normalized such that ${\bf J}\cdot{\bf J}=1$. The classical evolution of ${\bf J}$ is governed by the equations
\[
\dot{J_i}=\{J_i,H\},\quad \{J_i,J_j\}=\epsilon_{ijk}J_k
\]
where $\{\cdot\,,\cdot\}$ are the Poisson brackets. Their solution can be written 
as the mapping
\begin{equation}
{\bf J}^{n+1} = \left[\begin{array}{c} J_1^{n+1} \\ J_2^{n+1} \\ J_3^{n+1} \end{array}\right]  =  \left[\begin{array}{ccc} \cos{\omega} & -\sin{\omega} & 0 \\ \sin{\omega} & \cos{\omega} & 0 \\ 0 & 0 & 1 \end{array}\right]\left[\begin{array}{ccc} 1 & 0 & 0 \\ 0 & \cos{\mu s^n} & -\sin{\mu s^n} \\ 0 & \sin{\mu s^n} & \cos{\mu s^n} \end{array}\right]\left[\begin{array}{c} J_1^n \\ J_2^n \\ J_3^n\end{array}\right] \equiv \mbox{F}(s^n){\bf J}^n \label{map}
\end{equation}
where $s^n\equiv\sgn J_1^n$, which takes ${\bf J}$ from just before a kick to one 
period later. Here $\sgn x$ is the signum function with the convention $\sgn 0=0$. 

\begin{figure}[t]
\includegraphics[scale=0.75]{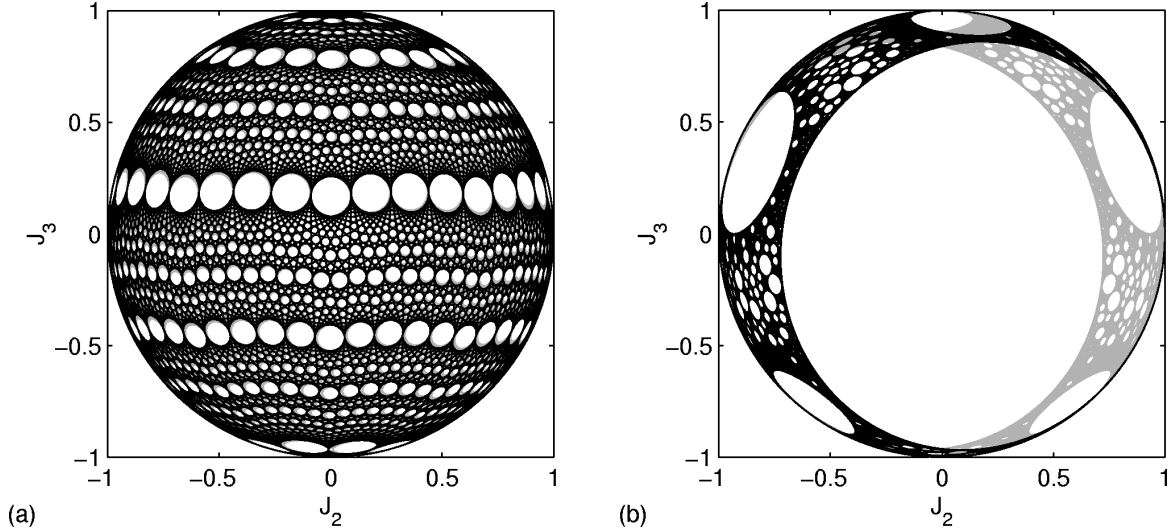}
\caption{The phase space of the classical map when $\omega=\pi(\sqrt{5}-1)$ and (a) $\mu=0.02$, (b) $\mu=\omega$.}
\label{fig1}
\end{figure}

The mapping is deceptively simple. It first rotates the eastern hemisphere ($J_1>0$) 
through an angle $\mu$ and the western hemisphere ($J_1<0$) through an angle 
$-\mu$. The great circle $J_1=0$ remains fixed. Then the entire sphere is rotated 
about the $J_3$ axis through an angle $\omega$. Thus the map rotates every point 
on the sphere in a piecewise linear fashion except on $J_1=0$ where its Jacobian is singular. 
Surprisingly, this leads to a phase space of great structural complexity. In Figure \ref{fig1} 
we give examples for two different parameter values. The eastern hemisphere is shown in 
black, the western in gray. We will only consider the semiclassical quantization 
when $\mu=\omega=\pi(\sqrt{5}-1)$ as in Figure \ref{fig1}(b).

Note that if an orbit of our map does not have a point on $J_1=0$, then its 
Lyapunov exponents are zero, making it stable. If however, it does have a point on 
$J_1=0$, then its Lyapunov exponents are undefined. We overcome this problem by 
simply defining an unstable orbit to be one with a point on $J_1=0$, and all 
others stable. The unstable set \footnote{It is not known whether the unstable set has zero measure. We 
currently believe that it has positive measure but empty interior. This would 
mean that the set of circular resonances is dense in phase space. See \cite{scott} 
for more on this topic.} is defined to be the closure of the set of all 
images and preimages of the great circle $J_1=0$. In figures \ref{fig1} and \ref{fig2} the unstable 
set is in black and gray. The circular holes are resonances consisting entirely of 
stable orbits. At their center lies a stable periodic orbit. It is these stable 
periodic orbits which will be used for the semiclassical quantization. All other 
points inside the resonances rotate about the central periodic orbit in a 
linear fashion. The period of rotation is an irrational multiple of $\pi$ 
for almost all values of the parameters $\mu$ and $\omega$, and hence, no other 
periodic orbits are located inside the resonances. In our case of study 
(Figure \ref{fig1}(b)), we have chosen $\mu$ to be large enough such that no 
island chains of first-order resonances form to create periodic quasi-tori 
(as in Figure \ref{fig1}(a)). Hence we have done our best to ensure that all 
of the stable periodic orbits are sufficiently isolated. The unstable periodic 
orbits are shown not to affect leading terms in the asymptotic analysis.  

We have previously shown \cite{scott} that it is possible to label every point of a stable \footnote{The unstable periodic 
orbits can be included by allowing 0's in the sequence. It can be shown that at 
most two zeros can occur in a sequence. Hence every unstable periodic orbit has 
exactly one or two points on the symmetry line $J_1=0$. See \cite{scott} for details.} periodic orbit of least 
period $n$ with a unique sequence $\{s^k=\pm 1\}_{k=1..n}$. The position of the 
point is given by the solution of
\begin{equation}
{\bf J}=\mbox{F}(s^n)\mbox{F}(s^{n-1})\dots \mbox{F}(s^1){\bf J}\equiv\mbox{R}{\bf J}
\label{per}
\end{equation}
lying on the unit sphere with $\sgn J_1=s^1$. The other $n-1$ points of the 
periodic orbit are found by substituting the $n-1$ cycles of the original sequence 
into (\ref{per}). Hence, if $\{{\bf J}^k\}_{k=1..n}$ is a stable periodic orbit 
then the sequence corresponding to its first point is 
$\{\!\sgn J_1^k\}_{k=1..n}$. Note that the (unnormalized) solution of (\ref{per}) is simply the 
axis of rotation of R, namely
\[
{\bf J}=(\mbox{R}_{23}-\mbox{R}_{32}\, ,\; \mbox{R}_{31}-\mbox{R}_{13}\, ,\; \mbox{R}_{12}-\mbox{R}_{21}). 
\]

Although every periodic point is uniquely represented by a sequence, not every 
sequence represents a periodic point. Hence we still need to find which sequences 
are legitimate. This may be done by checking each of the $2^n$ different possible 
sequences for a period-$n$ orbit. However, this is computationally expensive for 
large periods. An alternative is to exploit symmetries of our mapping. We have 
previously conjectured \cite{scott} that every stable periodic orbit has exactly one or two 
points on one of the following symmetry lines on the unit sphere: 
\begin{eqnarray}
J_1 &=& -J_2\tan{\omega/2} \nonumber \\
J_2 &=& J_3\tan(\mu/2\sgn J_1) \nonumber \\
J_2 &=& J_1\tan{\omega/2}. \nonumber
\end{eqnarray}
It is easy to prove that if a stable periodic orbit has one point on a particular 
symmetry line, then it has no more than two points on this line. However, we were 
not able to show that a stable periodic orbit {\it must} have a point on one of 
these lines, though numerical investigations suggest they do. We have checked all 
$2^{41}-2$ of the possible sequences for a stable periodic orbit of period 
$\leq 40$ and only found those of the above type. If one assumes our conjecture to 
be true then it is computationally easy to find all of the stable periodic orbits 
with periods upwards of 300 000. This may be done by finding where each symmetry line 
intersects with its image; an effortless task to solve numerically by virtue of the 
simplicity of our mapping. Every image of a symmetry line is just a collection of 
arc segments on the sphere. Once a periodic point is found, the size of its 
circular resonance can be calculated by noting that the resonance of at least one 
point of the orbit must touch $J_1=0$. This is because the boundary of each 
resonance forms part of the unstable set, and hence, iterates arbitrarily close 
to $J_1=0$. The sections of the symmetry line which intersect this periodic 
orbit's resonances contain no other periodic orbits and may be ignored for the 
remainder of our search. 

Following Ku\'s et al. \cite{kus} we rewrite the map in terms of the complex 
variables $\gamma$ and $\gamma^*$
\begin{equation}
\gamma_{n+1}^{\phantom{*}}=e^{i\omega}\frac{\gamma_n^{\phantom{*}}-is^n\tan{\frac{\mu}{2}}}{1-i\gamma_n^{\phantom{*}}s^n\tan{\frac{\mu}{2}}}\equiv\Gamma(\gamma_n^{\phantom{*}},\gamma_n^*) \label{cmap}
\end{equation}
where
\[
\gamma\equiv\frac{J_1+iJ_2}{1+J_3}=e^{i\phi}\tan\frac{\theta}{2}
\]
is the stereographic projection of the unit sphere 
${\bf J}=(\sin\theta\cos\phi,\sin\theta\sin\phi,\cos\theta)$ onto the complex 
plane. We then construct its generating function ($\re\gamma\neq 0$)
\begin{equation}
W(\Gamma^*,\gamma)=\log\left[(1+\gamma)(1+e^{i\omega}\Gamma^*)+e^{i\mu s}(1-\gamma)(1-e^{i\omega}\Gamma^*)\right]+C \label{W}
\end{equation}
where $s=\sgn\!\re\gamma$ \footnote{There is some ambiguity in the choice of $s$, e.g. (\ref{s2}) is also valid. One requires $s=\sgn\!\re\gamma$ in a neighborhood of $(\Gamma(\gamma)^*,\gamma)$ (guaranteed by the uniqueness of $W$ in this neighborhood).} 
and $C$ is an arbitrary constant. The mapping $\Gamma(\gamma,\gamma^*)$ can be 
rederived via the relations
\begin{equation}
\frac{\partial W}{{\partial\Gamma}^*}=\frac{\Gamma}{1+{\Gamma\Gamma}^*} \qquad \frac{\partial W}{\partial\gamma}=\frac{\gamma^*}{1+\gamma\gamma^*}. \label{dW}
\end{equation} 
It has been shown \cite{kus} that for nearly any area preserving map 
$\Gamma(\gamma,\gamma^*)$ it is possible to find a generating function 
$W(\Gamma^*,\gamma)$ with $\gamma$ and $\Gamma^*$ as its independent variables 
with the above relations (\ref{dW}). Except at the discontinuity $\re\gamma=0$, our particular map (\ref{cmap}) is 
locally only a function of $\gamma$, $\Gamma=\Gamma(\gamma)$. Hence we also have 
the relations
\begin{equation}
\frac{\partial^2 W}{{{\partial\Gamma}^*}^2}=\frac{-\Gamma^2}{(1+{\Gamma\Gamma}^*)^2} \qquad \frac{\partial^2 W}{\partial\gamma^2}=\frac{-{\gamma^*}^2}{(1+\gamma\gamma^*)^2}. \label{d2W}
\end{equation}

\section{Quantum map}\label{sectionIII}

The solution of Schr\"{o}dinger's equation with the Hamiltonian (\ref{H}) can be 
written in terms of a quantum map
\[
|\psi_{n+1}\rangle=e^{-i\omega\hat{J}_3}e^{-i\mu|\hat{J}_1|}|\psi_n\rangle\equiv\hat{F}|\psi_n\rangle
\]
where $\hat{F}$ is a Floquet operator and the angular momentum operators 
$\hat{J}_i$ satisfy 
\[
[\hat{J}_i,\hat{J}_j]=\epsilon_{ijk}\hat{J}_k, \qquad \hat{{\bf J}}^2=j(j+1) \qquad j=0,\textstyle\frac{1}{2},1,\textstyle\frac{3}{2},\dots
\]
For simplicity we have replaced $\hat{{\bf J}}$ with $\hbar\hat{{\bf J}}$ and 
noted that the normalization condition is now $\hbar^2\hat{{\bf J}}^2=1$ i.e. 
$\hbar=1/\sqrt{j(j+1)}$. The classical limit is now approached by letting 
$j\rightarrow\infty$. The quantum state $|\psi^n\rangle$ is a member of a Hilbert 
space spanned by the $2j+1$ orthonormal eigenstates of $\hat{J}_3$
\begin{eqnarray}
\hat{J}_3|j,m\rangle &=& m|j,m\rangle \nonumber\\
(\hat{J}_1\pm i\hat{J}_2)|j,m\rangle &=& \sqrt{(j\mp m)(j\pm m+1)}|j,m\pm 1\rangle \nonumber
\end{eqnarray}
where $-j\leq m\leq j$. By rewriting the Floquet operator as 
\begin{equation}
\hat{F}=e^{-i\omega\hat{J}_3}e^{-i\frac{\pi}{2}\hat{J}_2}e^{-i\mu|\hat{J}_3|}e^{i\frac{\pi}{2}\hat{J}_2} \label{F2}
\end{equation}
where
\[
|\hat{J}_3|=\sum_{m=-j}^j|m||j,m\rangle\langle j,m|
\]
one can find the matrix elements $\langle j,m|\hat{F}|j,n\rangle$ and hence, the 
quantum traces $\tr\hat{F}^n$.

The nonorthogonal overcomplete set of spin coherent 
states \cite{perelomov} make a more suitable basis for the semiclassical analysis. 
They are parametrized by the stereographic projection variable $\gamma$ and can be 
defined as a rotation of the minimum uncertainty eigenstate $|j,j\rangle$ 
\footnote{Note that this differs from the more common choice $|j,-j\rangle$.}
\begin{eqnarray}
|\gamma\rangle &\equiv& e^{i\theta(\hat{J}_1\sin\phi-\hat{J}_2\cos\phi)}|j,j\rangle \nonumber\\
&=& (1+\gamma\gamma^*)^{-j}\sum_{m=-j}^j\left(\frac{(2j)!}{(j+m)!(j-m)!}\right)^\frac{1}{2}\gamma^{j+m}|j,m\rangle \nonumber
\end{eqnarray}
so that
\[
\langle\gamma|\hat{{\bf J}}|\gamma\rangle=j(\sin\theta\cos\phi,\sin\theta\sin\phi,\cos\theta).
\]
The spin coherent states have minimal dispersion 
\[
\langle\gamma|\Delta\hat{{\bf J}}^2|\gamma\rangle=\langle\Delta\hat{{\bf J}}^2\rangle_{\text{min}}=j
\]
and are thus closest to classical states. The trace of an operator in this basis 
\begin{equation}
\tr\hat{A}=\frac{2j+1}{\pi}\int\frac{d^2\gamma}{(1+\gamma\gamma^*)^2}\langle\gamma|\hat{A}|\gamma\rangle \label{trace}
\end{equation}
results from the following resolution of unity
\begin{equation}
\frac{2j+1}{\pi}\int\frac{d^2\gamma}{(1+\gamma\gamma^*)^2}|\gamma\rangle\langle\gamma|=1 \label{resol}
\end{equation}
where $d^2\gamma=d(\re\gamma)d(\im\gamma)$.

\section{Semiclassical matrix elements}\label{sectionIV}

The semiclassical matrix element $\langle\Gamma|\hat{F}|\gamma\rangle_{\text{sc}}$ 
is the leading term in the asymptotic expansion ($j\rightarrow\infty$) of the 
exact quantum matrix element. To proceed with its derivation, first consider the 
matrix element
\begin{eqnarray}
\langle\xi|e^{-i\mu|\hat{J}_3|}|\eta\rangle = \frac{(\xi^*\eta)^j}{(1+\xi\xi^*)^j(1+\eta\eta^*)^j}&&\sum_{m=-j}^j \frac{(2j)!}{(j+m)!(j-m)!}(\xi^*\eta)^m e^{-i\mu|m|} \nonumber\\
= \frac{(\xi^*\eta)^j}{(1+\xi\xi^*)^j(1+\eta\eta^*)^j}&&\frac{(2j)!}{(j!)^2}\Bigg[-1+\sum_{m=0}^j \frac{(j!)^2}{(j+m)!(j-m)!}(\xi^*\eta e^{-i\mu})^m \nonumber \\
&&\qquad\qquad\quad+\sum_{m=0}^j \frac{(j!)^2}{(j+m)!(j-m)!}(\xi^*\eta e^{i\mu})^{-m}\Bigg] \nonumber\\
=\frac{(\xi^*\eta)^j}{(1+\xi\xi^*)^j(1+\eta\eta^*)^j}&&\frac{(2j)!}{(j!)^2}\Bigg[-1+{\,}_2F_1\Big(-j,1;j+1;-\xi^*\eta e^{-i\mu}\Big)
 \nonumber \\
&&\qquad\qquad\quad+{\,\,}_2F_1\Bigg(-j,1;j+1;-\frac{e^{-i\mu}}{\xi^*\eta}\Bigg)\Bigg] \nonumber
\end{eqnarray}
where $|\xi\rangle$ and $|\eta\rangle$ are arbitrary coherent states, and ${}_2F_1(-j,1;j+1;z)$ is a hypergeometric function. In Appendix \ref{appendixA} 
we derive an asymptotic expansion of this function when $j$ is large. Using these 
results one obtains 
\begin{eqnarray}
\langle\xi|&&e^{-i\mu|\hat{J}_3|}|\eta\rangle=\frac{e^{-ij\mu s}(1+e^{i\mu s}\xi^*\eta)^{2j}}{(1+\xi\xi^*)^j(1+\eta\eta^*)^j}+\frac{(\xi^*\eta)^j}{(1+\xi\xi^*)^j(1+\eta\eta^*)^j}\Bigg\{\sum_{m=0}^{n-1}\frac{\Gamma(m+1/2)(2j)!}{2\sqrt{\pi}j!(j+m)!}\cdot\nonumber\\
&&\cdot\left[\frac{1+\xi^*\eta e^{-i\mu}}{1-\xi^*\eta e^{-i\mu}}\left(\frac{-4\xi^*\eta e^{-i\mu}}{(1-\xi^*\eta e^{-i\mu})^2}\right)^m-\frac{1+\xi^*\eta e^{i\mu}}{1-\xi^*\eta e^{i\mu}}\left(\frac{-4\xi^*\eta e^{i\mu}}{(1-\xi^*\eta e^{i\mu})^2}\right)^m\right]+O(j^{-n})\Bigg\} \label{scmatrix}
\end{eqnarray}
if $|\xi^*\eta|\neq 1$, where $s=\sgn(1-|\xi^*\eta|)$. This approximation is 
exponentially small for large $j$ except when $\xi=e^{i\mu s}\eta$, in which case 
$s=\sgn(1-|\eta|)$ and hence $\xi$ is the image of $\eta$ under the classical map 
corresponding to $e^{-i\mu|\hat{J}_3|}$. When this happens the first term in our 
expansion is $O(1)$ and exponentially dominates. Hence we take
\begin{equation}
\langle\xi|e^{-i\mu|\hat{J}_3|}|\eta\rangle_{sc}= \frac{e^{-ij\mu s}(1+e^{i\mu s}\xi^*\eta)^{2j}}{(1+\xi\xi^*)^j(1+\eta\eta^*)^j} \label{scJ3}
\end{equation}
as the semiclassical approximation of the quantum matrix element. In the case 
$|\xi^*\eta|=1$ one finds that the leading term in the asymptotic expansion is 
$O(1)$ only when $\xi=e^{\pm i\mu}\eta$ with $|\xi|=|\eta|=1$. However the 
classical map $\xi=e^{i\mu s}\eta$ does not behave in this manner on the singular 
line $|\eta|=1$. In the next section we find that it is only when the mapping 
$\eta\rightarrow\xi$ forms, through (\ref{F2}), part of a classical periodic orbit 
that contributions to the semiclassical trace become important. Hence we may 
assume $|\xi^*\eta|\neq 1$, which amounts to ignoring the unstable set of our 
original classical map (\ref{map}).    

Now using (\ref{scJ3}) and the relations 
\begin{eqnarray*}
e^{i\frac{\pi}{2}\hat{J}_2}|\gamma\rangle&=&\left(\frac{1+\gamma}{1+\gamma^*}\right)^j\left|\frac{\gamma-1}{\gamma+1}\right\rangle \\
e^{i\omega\hat{J}_3}|\gamma\rangle&=&e^{i\omega j}|e^{-i\omega}\gamma\rangle
\end{eqnarray*}
with (\ref{F2}) one finds that
\begin{equation}
\langle\Gamma|\hat{F}|\gamma\rangle_{sc}= \frac{4^{-j}e^{-i\omega j}e^{-i\mu j}}{(1+{\Gamma\Gamma}^*)^j(1+\gamma\gamma^*)^j}\left[(1+\gamma)(1+e^{i\omega}\Gamma^*)+e^{i\mu s}(1-\gamma)(1-e^{i\omega}\Gamma^*)\right]^{2j} 
\end{equation}
where 
\begin{equation}
s=\sgn\left(1-\left|\frac{e^{-iw}\Gamma-1}{e^{-iw}\Gamma+1}\right|\cdot\left|\frac{\gamma-1}{\gamma+1}\right|\right). \label{s2}
\end{equation}
In particular, if $\Gamma=\Gamma(\gamma,\gamma^*)$ under the classical map 
(\ref{cmap}) then $s=\sgn\!\re\gamma$. The semiclassical matrix element can be 
rewritten in the two forms
\begin{eqnarray}
\langle\Gamma|\hat{F}|\gamma\rangle_{sc} &=& \left(\frac{\partial^2 W}{{\partial\Gamma}^*\partial\gamma}\right)^{\frac{1}{2}}\exp\Big[(2j+1)W(\Gamma^*,\gamma)\Big](1+{\Gamma\Gamma}^*)^{-j}(1+\gamma\gamma^*)^{-j} \label{Fsc1}\\
&=& \exp\Big[2j\,W(\Gamma^*,\gamma)\Big](1+{\Gamma\Gamma}^*)^{-j}(1+\gamma\gamma^*)^{-j} \label{Fsc2}
\end{eqnarray}
by putting $C=-\log{2}-i\omega/2-i\mu/2$ in (\ref{W}). The first (\ref{Fsc1}) is 
reminiscent of that for the kicked top \cite{kus} while the second (\ref{Fsc2}) 
results from
\begin{equation}
\frac{\partial^2 W}{{\partial\Gamma}^*\partial\gamma}=\exp\Big[-2W(\Gamma^*,\gamma)\Big] \label{d2W2}
\end{equation}
which is special to this mapping. In both cases the semiclassical matrix element 
is fully determined by the classical generating function.

\section{Semiclassical traces and eigenphases}\label{sectionV}

We now wish to derive the semiclassical approximation to the trace
\begin{eqnarray}
\tr\hat{F}^n &=& \frac{2j+1}{\pi}\int\frac{d^2\gamma_1^{\phantom{*}}}{(1+\gamma_1^{\phantom{*}}\gamma_1^*)^2}\langle\gamma_1^{\phantom{*}}|\hat{F}^n|\gamma_1^{\phantom{*}}\rangle \label{traceg}\\
&=& \left(\frac{2j+1}{\pi}\right)^n\int\prod_{k=1}^n\frac{d^2\gamma_k^{\phantom{*}}}{(1+\gamma_k^{\phantom{*}}\gamma_k^*)^2}\langle\gamma_{k+1}^{\phantom{*}}|\hat{F}|\gamma_k^{\phantom{*}}\rangle \qquad\qquad (n+1\equiv 1) \nonumber\\
&\approx& \left(\frac{2j+1}{\pi}\right)^n\int\prod_{k=1}^n\frac{d^2\gamma_k^{\phantom{*}}}{(1+\gamma_k^{\phantom{*}}\gamma_k^*)^2}\exp\left[2j\left(W(\gamma_{k+1}^*,\gamma_k^{\phantom{*}})-\log(1+\gamma_k^{\phantom{*}}\gamma_k^*)\right)\right] \label{gint}
\end{eqnarray}
where we have used (\ref{trace}), (\ref{resol}) and the semiclassical matrix element (\ref{Fsc2}). In the asymptotic limit $j\rightarrow\infty$ this multiple integral is dominated by the contributions at saddle points where
\[
\frac{\partial W(\gamma_{k+1}^*,\gamma_k^{\phantom{*}})}{\partial\gamma_{k+1}^*}=\frac{\gamma_{k+1}^{\phantom{*}}}{1+\gamma_{k+1}^{\phantom{*}}\gamma_{k+1}^*} \qquad \frac{\partial W(\gamma_{k+1}^*,\gamma_k^{\phantom{*}})}{\partial\gamma_k^{\phantom{*}}}=\frac{\gamma_k^*}{1+\gamma_k^{\phantom{*}}\gamma_k^*} 
\]
for $k=1\dots n$ $(n+1\equiv 1)$. Recalling relations (\ref{dW}) we see that the 
periodic points of the classical map are saddle points. However these are 
not the only saddle points. Contributions from `ghost orbits' \cite{kus2} 
also affect the above integral. Whenever a periodic orbit is destroyed at 
a bifurcation point there is still a residual contribution to the integral. 
These contributions are not semiclassical, being exponentially smaller, 
and will be ignored at this stage. However, near the bifurcation point the semiclassical approximation will become 
inaccurate if $j$ is not large enough. By expanding the argument of the 
exponential up to second order one may approximate the contribution from 
each of the classical periodic points
\begin{eqnarray}
\tr\hat{F}^n &\approx& \sum_{\bm{\rho}}\left(\frac{2j+1}{\pi}\right)^n\int\prod_{k=1}^n\frac{d^2\gamma_k^{\phantom{*}}}{(1+\rho_k^{\phantom{*}}\rho_k^*)^2}\exp\left[2j\left(W(\rho_{k+1}^*,\rho_k^{\phantom{*}})-\log(1+\rho_k^{\phantom{*}}\rho_k^*)\right)\right] \nonumber\\
&&\quad\qquad\qquad\qquad\cdot\exp\left[2j\left(\exp\Big[-2W(\rho_{k+1}^*,\rho_k^{\phantom{*}})\Big]\gamma_{k+1}^*\gamma_k^{\phantom{*}}-\frac{1}{1+\rho_k^{\phantom{*}}\rho_k^*}\gamma_k^*\gamma_k^{\phantom{*}}\right)\right] \label{gauss}
\end{eqnarray}
where $\bm{\rho}=\{\rho_k\}_{k=1..n}$ is the periodic orbit corresponding to the 
periodic point $\rho_1$, and we have replaced $\bm{\gamma}$ by 
$\bm{\gamma}+\bm{\rho}$ and used (\ref{d2W}) and (\ref{d2W2}) for simplification. 
This multiple integral may now be solved analytically (see Appendix \ref{appendixB}). 
The result is our semiclassical approximation for the trace  
\begin{equation}
(\tr\hat{F}^n)_{sc}=\sum_{\bm{\rho}}\frac{\exp\Big[i(2j+1)S(\bm{\rho})\Big]}{2i\sin\Big[S(\bm{\rho})\Big]} \label{sctr}
\end{equation}
where the sum is taken over all periodic points (all unique points of each periodic orbit with period dividing $n$) and 
\[
iS(\bm{\rho})\equiv\sum_{k=1}^n W(\rho_{k+1}^*,\rho_k^{\phantom{*}})-\log(1+\rho_k^{\phantom{*}}\rho_k^*).
\]
$S$ is the action of the classical orbit \cite{kus}.

Consider the contribution to the semiclassical trace formula from a single 
periodic orbit $\bm{\rho}=\{\rho_k\}_{k=1..p}$ with least period $p$ and action 
$S(\bm{\rho})$ from a single traversal. This orbit will contribute when $n=kp$
\begin{eqnarray*}
(\tr\hat{F}^{kp})_{sc} &=& p\frac{\exp\Big[i(2j+1)kS(\bm{\rho})\Big]}{2i\sin\Big[kS(\bm{\rho})\Big]}+(\text{contributions from other orbits}) \\
&=& p\sum_{m=0}^{\infty}\exp\Big[i(j-m)2kS(\bm{\rho})\Big]+(\text{contributions from other orbits}).
\end{eqnarray*}
However the exact quantum mechanical trace is
\[ 
\tr\hat{F}^{kp}=\sum_{l=0}^{2j}\exp(-ikp\phi_l)
\]
where $\phi_l$ are the eigenphases of the Floquet operator. Upon comparing these 
last two equations we see that for each $m\geq 0$ there must be $p$ distinct 
eigenphases with the property 
\[
p\phi_l \;\rightarrow\; 2(m-j)S(\bm{\rho})\;\;\;\text{mod}\;\; 2\pi \qquad\text{as}\qquad j\rightarrow\infty,
\]
and hence, we obtain the following semiclassical approximation to the eigenphases
\begin{equation}
(\phi_{mnp})_{sc}=\frac{2(m-j)}{p}S(\bm{\rho})+\frac{2n\pi}{p} \;\;\;\text{mod}\;\; 2\pi \qquad n=1,..\,p\qquad m=0,1,\dots \label{eigsc}
\end{equation}
This approximation is not surprising. In the classical phase space each stable 
periodic point is at the center of a circular resonance, and all orbits in this 
resonance rotate about the periodic orbit in a linear fashion. Hence the map 
$\hat{F}^p$ behaves, to a first-order approximation, like the linear top 
$\exp[2iS(\bm{\rho})\hat{J}_3]$ near the periodic point. The point $J_3=-1$ of 
the linear top corresponds to the periodic point. When $m=0$, the $p$ different 
eigenphases correspond to the $p$ different ground states associated with the 
periodic orbit. These ground states are localized on the periodic orbit and are 
$p$-fold `cat states' \cite{janszky} (see Eq. (\ref{cat}) and Fig. \ref{fig7}(a)). The higher order states are ranked via $m$ and form rings 
around the classical periodic orbit similar to that of the linear top (see Figs. \ref{fig7}(b) and (c)). However, 
if $j$ is finite, the semiclassical eigenphases will only be good approximates to 
the quantum for $m\lesssim M$, for some $M$, since only finitely many will `fit' in 
the classical circular resonance. We can estimate $M$ by considering the 
eigenstates of the linear top. These eigenstates have circular Husimi functions 
$|\langle\gamma |j,l\rangle |^2$ localized at an angle $\theta$ satisfying 
\[
\tan^2(\theta/2)=\frac{j+l}{j-l} \qquad -j\leq l\leq j.
\]
If we define $\theta_r$ to be the angular separation between a point of the 
periodic orbit and the boundary of its circular resonance, and find the 
eigenstate $l_r$ of the linear top which is localized at this angle, then we 
obtain an estimate for $M$ 
\begin{equation}
M(\bm{\rho})=j+l_r=2j\sin^2(\theta_r/2)=j-j\sqrt{1-R(\bm{\rho})^2} \label{M}
\end{equation}
where 
\[
R(\bm{\rho})\equiv\min_{k=1..p} |J_1^k|\quad (=\sin\theta_r).
\]
The calculation of $R(\bm{\rho})$ follows from the fact that at least one of 
the circular resonances associated with the periodic orbit will be tangent to 
the great circle $J_1=0$.

\section{Numerical investigations of the semiclassical traces}\label{sectionVI}
 
\begin{figure}[t]
\includegraphics[scale=0.9]{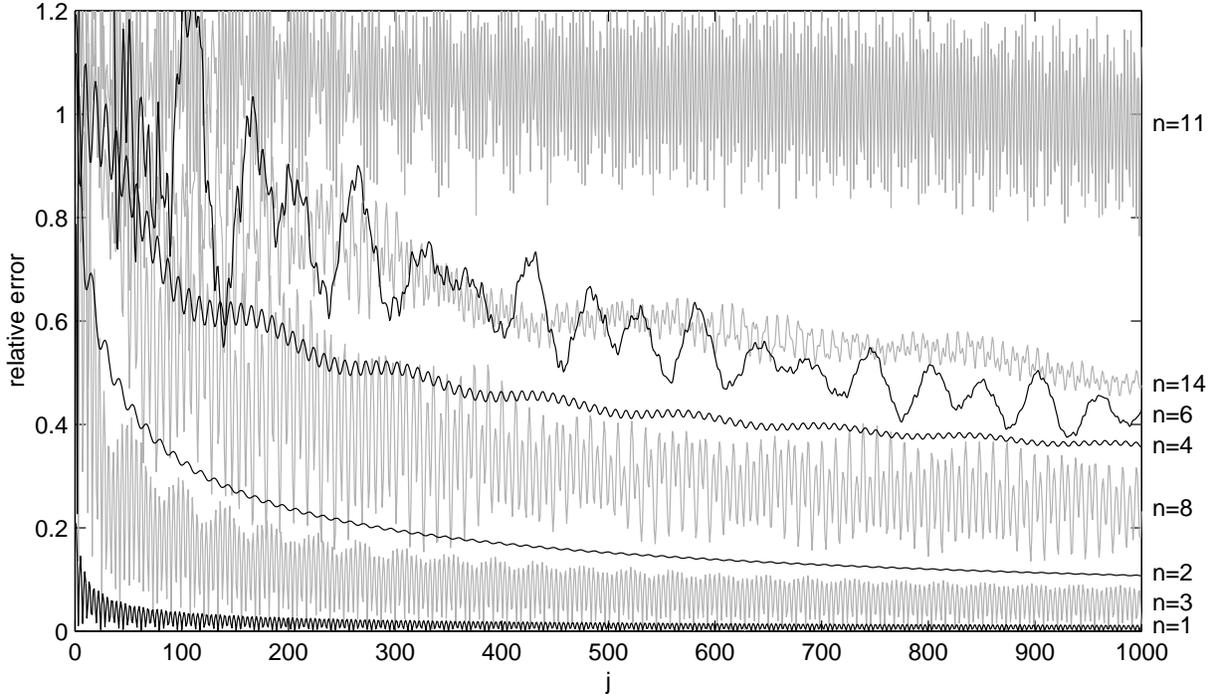}
\caption{The relative error (\ref{error}) of various semiclassical traces versus $j$ for $\mu=\omega=\pi(\sqrt{5}-1)$.}
\label{fig2}
\end{figure}

\begin{figure}[p]
\includegraphics[scale=0.81]{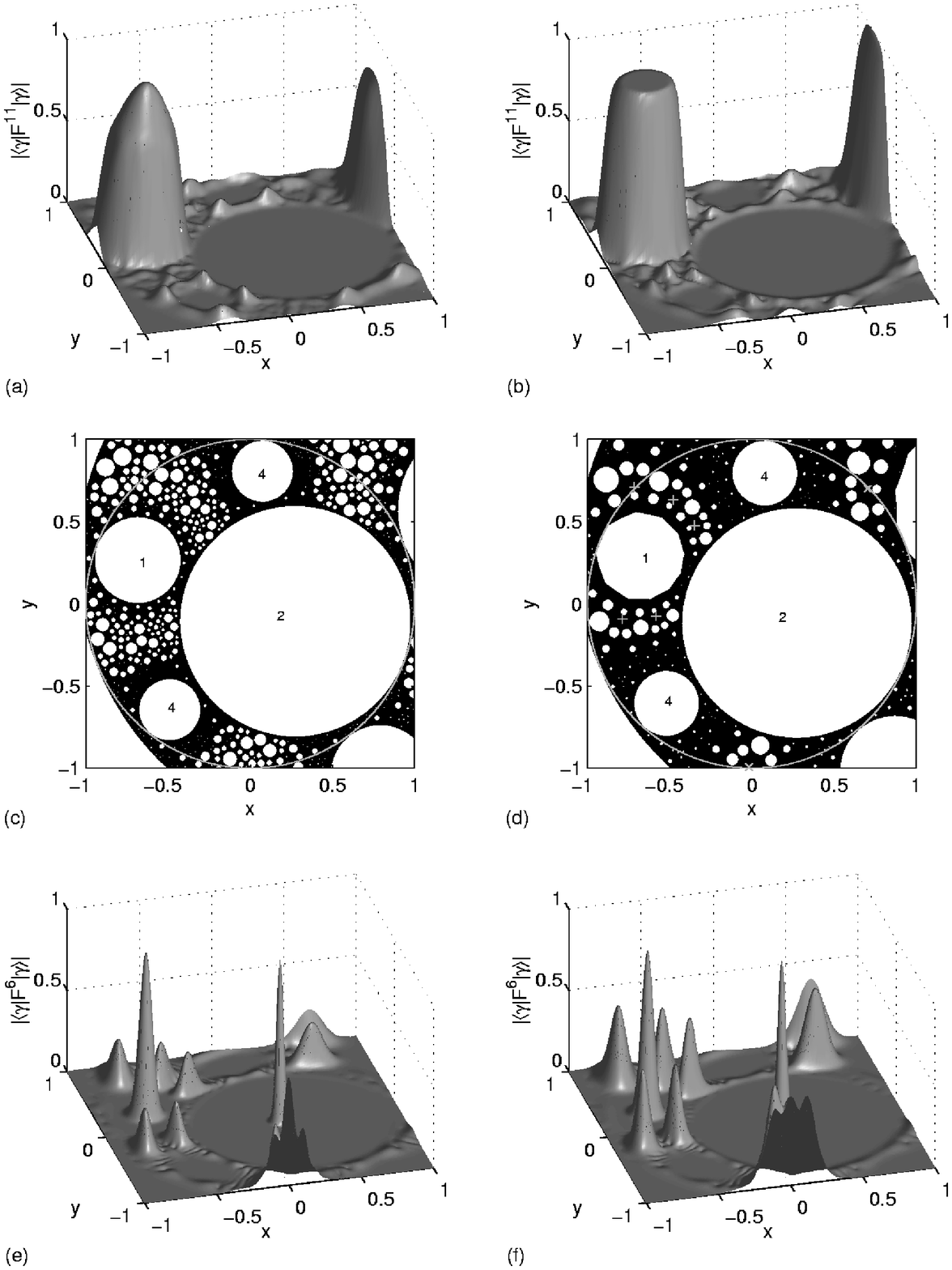}
\caption{The quantum diagonal matrix element 
$\langle\gamma|\hat{F}^n|\gamma\rangle$ when $j=300$ using the 
stereographic projection (\ref{xy}), and with (a) $n=11$, $\mu=\omega$, 
(b) $n=11$, $\mu=\mu_1\approx\omega+0.06556$, (e) $n=6$, $\mu=\omega$, and
(f) $n=6$, $\mu=\mu_1$. The stereographic projection of the classical phase 
space when (c) $\mu=\omega$ and (d) $\mu=\mu_1$.}
\label{fig3}
\end{figure}

\begin{table}[b]
\caption{Actions of all stable periodic orbits with period $p\leq 25$ when $\mu=\omega=\pi(\sqrt{5}-1)$.}
\label{table1}
\begin{ruledtabular}
\begin{tabular}{ccc|ccc|cc}
$p$ & $S$ && $p$ & $S$ && $p$ & $S$ \\ \hline
1 & -4.8441 && 14 & -35.5088 && 20 & -26.4359\\
1 & 1.4391 && 14 & -10.3761 && 20 & -39.0023\\
2 & -0.6890 && 16 & -20.0022 && 20 & -64.5127\\
4 & -6.4608 && 16 & -45.1734 && 20 & -1.6809\\
9 & -17.7025 && 16 & -7.4743 && 22 & -48.6527\\
9 & -11.2193 && 18 & -54.8424 && 22 & -23.5199\\
12 & -13.2882 && 18 & -4.5769 && 24 & -32.8587\\
\end{tabular}
\end{ruledtabular}
\end{table}
The semiclassical approximation to the traces (\ref{sctr}) is exact in the limit $j\rightarrow\infty$. But for finite $j$ we need to test how large $j$ must be before this approximation is an accurate one. We will only consider the case when $\mu=\omega=\pi(\sqrt{5}-1)$. The actions $S$ of all stable periodic orbits with period $p\leq 25$ have been compiled in Table \ref{table1}. Figure \ref{fig2} shows the relative error of various semiclassical traces versus $j$ 
\begin{equation}
\text{relative error}=\frac{\left|(\tr\hat{F}^n)_{sc}-\tr\hat{F}^n\right|} {\sqrt{\overline{|(\tr\hat{F}^n)_{sc}|^2}}} \label{error}
\end{equation}
where
\[
\overline{|(\tr\hat{F}^n)_{sc}|^2}=\sum_{\bm{\rho}}\frac{1}{4\sin^2\Big[S(\bm{\rho})\Big]}.
\]
Note that some of the traces can be quite slow to converge. In the case of 
$n=11$ convergence is extremely slow. This was found to be caused by the 
failure of the semiclassical trace formula (\ref{sctr}) when 
$\sin S(\bm{\rho})=0$. If $n=11$ the only contributions to the trace are 
from two period-1 orbits each with action $S\text{ mod }2\pi=1.4391$ from 
a single traversal. Hence $\sin 11S=-0.1218$, which is close to zero 
making our Gaussian approximation (\ref{gauss}) an inaccurate one. Higher 
order terms from a complete asymptotic expansion of the integral 
(\ref{gint}) would need to be calculated for increased accuracy. The 
problem is further exposed when the quantum diagonal matrix element 
$\langle\gamma|\hat{F}^n|\gamma\rangle$ is plotted. This is done for 
$j=300$ in Figure \ref{fig3}(a) using the stereographic projection
\begin{equation}
x=\frac{J_2}{1+J_1} \qquad y=\frac{J_3}{1+J_1}. \label{xy}
\end{equation}
Figure \ref{fig3}(c) shows the corresponding classical phase space. The periodic points 
of period 1, 2 and 4 are labeled and the circle $J_1=0$ is shown in gray. One 
can see that an overwhelming contribution to the trace (\ref{traceg}) in the 
form of a large hump is produced by the presence of a classical period-1 orbit. 
However at $j=300$ the hump does not take the form of a sharp Gaussian peak and is 
instead quite wide. 

\begin{figure}[t]
\includegraphics[scale=0.92]{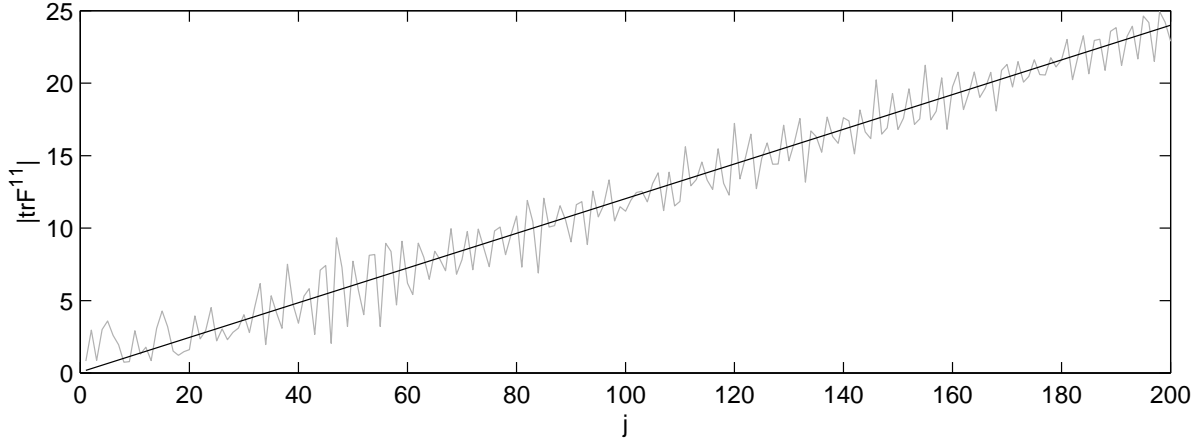}
\caption{The semiclassical (black) and quantum (gray) trace $\tr\hat{F}^{11}$ when $\mu=\mu_1$.}
\label{fig4}
\end{figure}

Complete failure of the semiclassical trace formula occurs when $\sin S(\bm{\rho})=0$. If this happens in the classical map the local rotation about the stable periodic point $\rho$, given by $2S(\bm{\rho})$, becomes a multiple of $2\pi$. In our case
this occurs when $\mu=\mu_1\approx\omega+0.06556$ where $\mu_1$ is defined via
\[
1+2\cos \textstyle\frac{10}{11}\pi = \cos\omega+\cos\omega\cos\mu_1+\cos\mu_1.
\]
The two period-1 orbits now have the action $S\text{ mod }2\pi=5\pi/11$. In more generic systems a period-11 orbit would bifurcate from each period-1 orbit at this point.
However no such bifurcation occurs in our mapping. Instead the circular resonance enclosing each of the period-1 orbits takes the form of an 11 sided polygon (Figure \ref{fig3}(d)). Every point inside this polygon rotates $10\pi/11$ radians about the period-1 orbit at each iteration of the mapping, and hence, is a period-11 orbit.
The contribution to the semiclassical trace in these special cases is
\[
(\tr\hat{F}^n)_{sc}=(2j+1)\sum_{\bm{\rho}}\frac{A(\rho)}{4\pi}
\]
where $A(\rho)$ is the area on the unit sphere covered by the polygon. In our case we have a regular polygon with area
\[
A=2\pi-2n\arcsin\left(\sqrt{1-{J_1}^2}\sin(\pi/n)\right)
\]
where $n=11$ and the period-1 orbit is located at $(J_1,J_2,J_3)$ with 
\[
J_1=\pm\cos(\omega/2)/\sqrt{1+\sin^2(\omega/2)\cot^2(\mu_1/2)}.
\]
In Figure \ref{fig4} we have plotted the semiclassical and quantum traces for this case.
The semiclassical displays a good first-order approximation to the 
quantum. Note that in the limit $j\rightarrow\infty$ the trace is infinite, 
whereas for $\mu=\mu_1\pm\epsilon$ ($\epsilon>0$) the trace limits to a large 
but finite number. The semiclassical trace {\it is} this limit in the latter case.
Consequently, when $\epsilon$ becomes small (but still nonzero), the 
semiclassical trace (\ref{sctr}) will be an extremely poor approximate for the 
quantum, since $j$ needs to be ever larger before the quantum trace 
reaches its semiclassical limit. Hence the poor convergence in Figure \ref{fig2} when 
$n=11$ is explained.

\begin{figure}[t]
\includegraphics[scale=0.92]{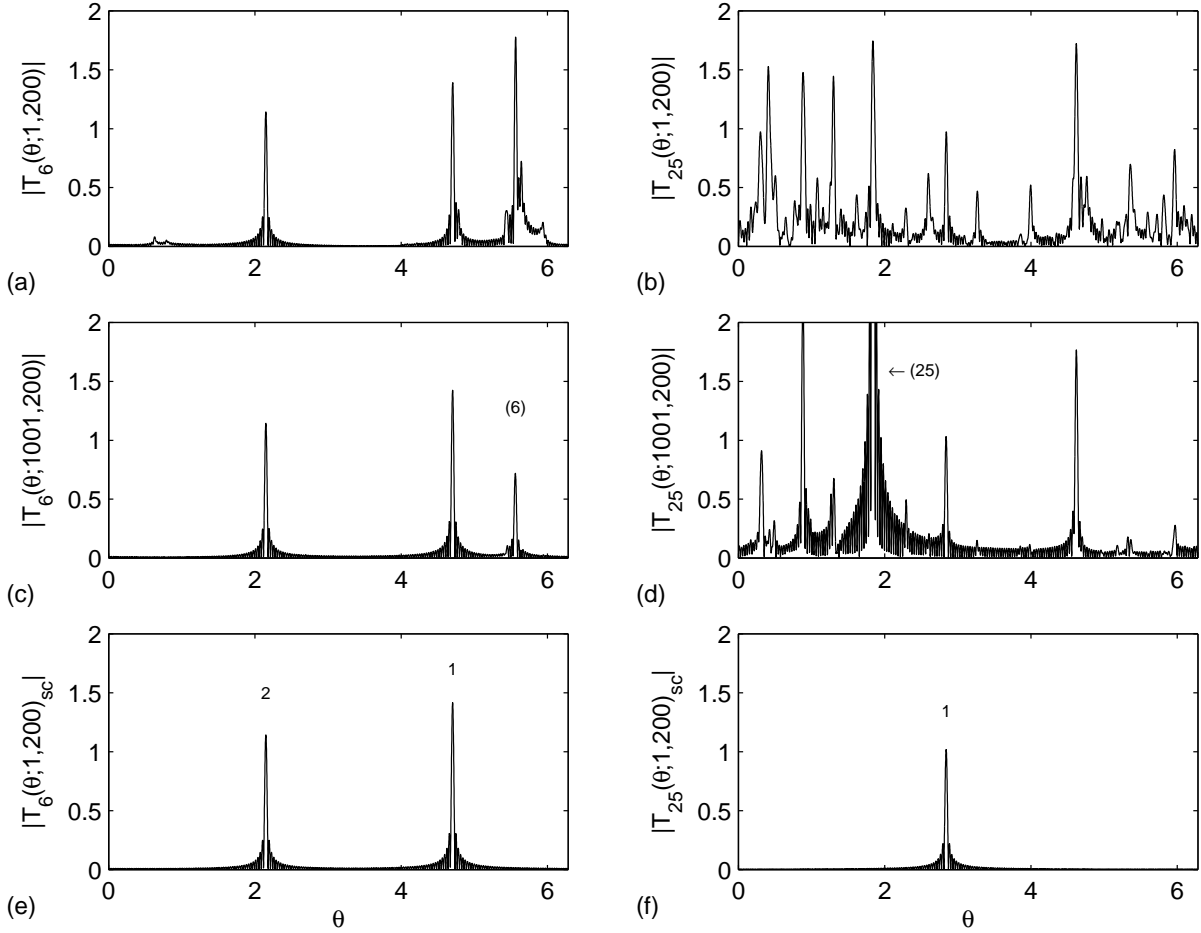}
\caption{The quantum Fourier transforms $T_n(\theta;j_0,200)$ when
(a) $n=6$, $j_0=1$, (b) $n=25$, $j_0=1$, (c) $n=6$, $j_0=1001$, and (d) 
$n=25$, $j_0=1001$. The semiclassical transforms when (e) $n=6$ and (f) 
$n=25$.}
\label{fig5}
\end{figure}
Another test of the semiclassical traces is to consider the quantum Fourier transforms 
\begin{equation}
T_n(\theta;j_0,N)=\frac{1}{N}\sum_{j=j_0}^{j_0+N}e^{-ij\theta}\tr\hat{F}^n \label{transform}
\end{equation}
which exhibit peaks at $\theta=2S(\bm{\rho})\text{ mod }2\pi$. In Figures 
\ref{fig5}(a) and (c) we have plotted the absolute value of the quantum transform $T_6$ when
$j_0=1$ and $j_0=1001$, respectively, and $N=200$. Figure \ref{fig5}(e) shows the 
transform of the semiclassical trace. The peaks have been labeled 
according to the classical periodic orbit from which they derive. Note 
that the quantum transforms exhibit a third peak with no semiclassical 
analogue. This peak illustrates the effect of `ghost orbits' \cite{kus2}.
It is the result of a residual contribution to the trace, left behind after 
a pair of period-6 orbits were destroyed when $\mu\approx\omega+0.05318$. 
In Figure \ref{fig3}(f) we have plotted the quantum diagonal matrix element 
$\langle\gamma|\hat{F}^6|\gamma\rangle$ when 
$\mu=\mu_1\approx\omega+0.06556$ and $j=300$. One finds sharp Gaussian 
peaks at the location of the classical periodic orbits of periods 1, 2 and 
6. The pair of period-6 orbits are marked by $+$'s and $\times$'s in the 
classical phase space (Figure \ref{fig3}(d)). In spite of there being no classical 
period-6 orbits when $\mu=\omega$, the quantum diagonal matrix element 
still displays ghost peaks at their previous locations (Figure \ref{fig3}(e)). 
These peaks, however, will vanish in the limit $j\rightarrow\infty$ unlike 
the case of $\mu=\mu_1$ where the peaks will instead increase to a height 
of unity. The exponential decrease in these ghost contributions is also 
exhibited in the quantum transforms (Figures \ref{fig5}(a) and (c)). Figures 
\ref{fig5}(b) and (d) show the quantum transform $T_{25}$. The only
contribution to the semiclassical trace is from the pair of period-1 
orbits (Figure \ref{fig5}(f)). But in the quantum transforms this has been swamped 
by other contributions. The largest peak in Figure \ref{fig5}(d) is the ghost of 
a pair of period-25 orbits which were destroyed at 
$\mu\approx\omega+0.0004288$. This peak has temporarily grown in size after 
increasing $j$. The other quantum peaks are also believed to be caused by 
residual quantum effects. However we are currently unable to give an exact 
reason for their presence.

The quantum transform $T_{25}$ paints a disheartening picture of exponentially 
proliferating quantum effects not accounted for in the simple semiclassical analysis. In general we found our semiclassical trace 
$(\tr\hat{F}^n)_{sc}$ to be plagued by errors when $n\gtrsim 10$. Although
our approximation becomes exact in the limit $j\rightarrow\infty$, and for
large $j$ the error is exponentially decreasing, this decay did not occur as 
quickly as we had hoped. In the next section we find that these inaccuracies 
make it impossible to extract semiclassical eigenphases from the traces.  

\section{Numerical investigations of the semiclassical spectrum}\label{sectionVII}

Consider the spectral density of the eigenphases
\begin{eqnarray*}
\rho(\theta;j) &\equiv& \sum_{m=0}^{2j}\sum_{k=-\infty}^{\infty}\delta(\theta-\phi_m+2k\pi) \\
&=& \frac{1}{2\pi}\sum_{m=0}^{2j}\sum_{n=-\infty}^{\infty}e^{in(\theta-\phi_m)} \\
&=& \frac{2j+1}{2\pi}+\frac{1}{\pi}\sum_{n=1}^{\infty}\re(e^{in\theta}\tr\hat{F}^n).
\end{eqnarray*}
Hence, by rewriting the spectral density in terms of traces of the Floquet
operator and using the semiclassical trace formula (\ref{sctr}), one may 
derive a semiclassical spectral density. However, this formula requires 
traces of the Floquet operator to all powers $n$, which invariably become 
inaccurate when $n$ is large. Thus, we instead consider the spectral 
density with limited resolution
\[
\rho(\theta;j,N)\equiv\frac{2j+1}{2\pi}+\frac{1}{\pi}\sum_{n=1}^N\re(e^{in\theta}\tr\hat{F}^n)
\]
\begin{figure}[t]
\includegraphics[scale=0.92]{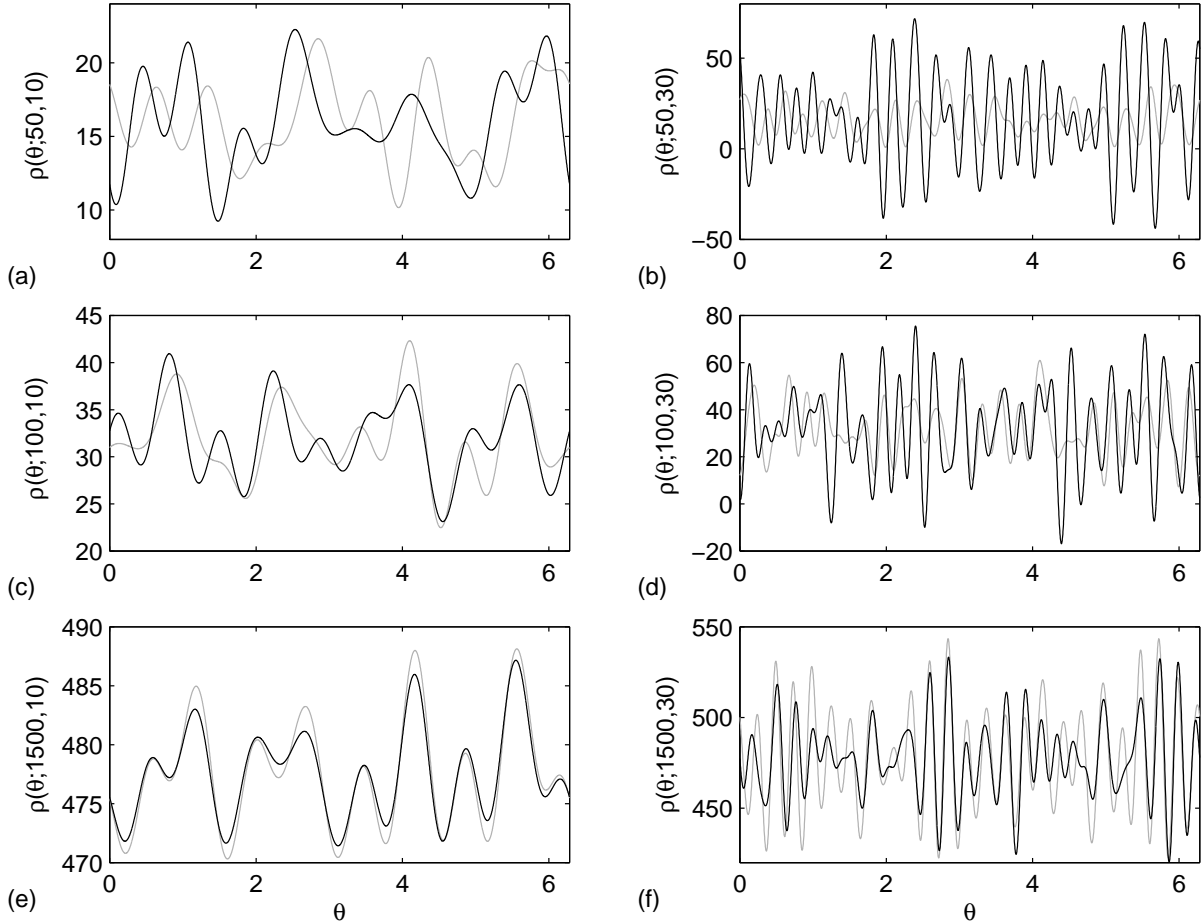}
\caption{The semiclassical (black) and quantum (gray) spectral density $\rho(\theta;j,N)$ when $N=10$ and (a) $j=50$, 
(c) $j=100$, (e) $j=1500$, and when $N=30$, (b) $j=50$, 
(d) $j=100$, (f) $j=1500$.}
\label{fig6}
\end{figure}

In figures \ref{fig6}(a), (c) and (e) we plot the semiclassical and quantum (gray) 
spectral densities when $N=10$ and $j=50$, $100$ and $1500$, respectively. 
Figures \ref{fig6}(b), (d) and (f) are the same except with $N=30$. The semiclassical 
approximation is of course most accurate when $j=1500$ and $N=10$. However one 
needs $N\gg 10$ to resolve the $3001$ eigenphases and even increasing
to $N=30$ one finds large discrepancies between the quantum and semiclassical 
spectral densities.

Another approach \cite{kus} is to use Newton's formulae \cite{macduffee} and rewrite 
the coefficients of the characteristic polynomial $\det(\hat{F}-z)$ in terms of
traces of the Floquet operator. All $2j+1$ eigenphases $\phi_m$ can then be extracted via
the roots $z_m=e^{-i\phi_m}$ of the characteristic polynomial. However we still need
accurate semiclassical approximates of the trace $\tr\hat{F}^n$ for 
$1\leq n\leq 2j+1$ (or $1\leq n\leq j+1$ if we enforce `selfinversiveness' on the 
polynomial i.e. $\det(\hat{F}-z)=z^{2j+1}\det\hat{F}\det(\hat{F}^\dag-z^{-1})$ where 
$\det\hat{F}=e^{-i\mu j(j+1)}$). The method of harmonic inversion \cite{main,main2} 
via filter-diagonalization \cite{wall,mandelshtam} also proved futile. 
The problem is not due to the extraction procedure but rather the inaccuracy of
our semiclassical trace $(\tr\hat{F}^n)_{sc}$ when $n$ becomes large. 
\begin{table}
\caption{Semiclassical and quantum eigenphases when $\mu=\omega=\pi(\sqrt{5}-1)$ and $j=100$.}
\label{table2}
\begin{ruledtabular}
\begin{tabular}{cccccc}
$m$ & $n$ & $p$ & $(\phi_{mnp})_{sc}$ & $(\phi)_{qu}$ & $\Delta$ \\ \hline
0 & 1 & 1 & 1.2064 & 1.2064 & $3\times 10^{-7}$\\
0 & 1 & $1'$ & 1.2064 & 1.2064 & $4\times 10^{-7}$\\
1 & 1 & 1 & 4.0846 & 4.0846 & $2\times 10^{-6}$\\
1 & 1 & $1'$ & 4.0846 & 4.0846 & $2\times 10^{-5}$\\
2 & 1 & 1 & 0.6797 & 0.6799 & $2\times 10^{-4}$\\
2 & 1 & $1'$ & 0.6797 & 0.6800 & $3\times 10^{-4}$\\
6 & 1 & 1 & 5.9093 & 5.8913 & $2\times 10^{-2}$\\
6 & 1 & $1'$ & 5.9093 & 5.8533 & $6\times 10^{-2}$\\
0 & 1 & 2 & 2.9306 & 2.9306 & $2\times 10^{-13}$\\
0 & 2 & 2 & 6.0722 & 6.0722 & $1\times 10^{-13}$\\
1 & 1 & 2 & 2.2416 & 2.2416 & $1\times 10^{-12}$\\
1 & 2 & 2 & 5.3832 & 5.3832 & $1\times 10^{-12}$\\
2 & 1 & 2 & 1.5525 & 1.5525 & $1\times 10^{-12}$\\
2 & 2 & 2 & 4.6941 & 4.6941 & $1\times 10^{-12}$\\
57 & 1 & 2 & 1.3544 & 1.3545 & $1\times 10^{-4}$\\
57 & 2 & 2 & 4.4960 & 4.4958 & $2\times 10^{-4}$\\
58 & 1 & 2 & 0.6654 & 0.6657 & $3\times 10^{-4}$\\
58 & 2 & 2 & 3.8070 & 3.8006 & $6\times 10^{-3}$\\
59 & 1 & 2 & 6.2595 & 6.2590 & $5\times 10^{-4}$\\
59 & 2 & 2 & 3.1179 & 3.1340 & $2\times 10^{-2}$\\
0 & 2 & 4 & 5.7403 & 5.7420 & $2\times 10^{-3}$\\
0 & 4 & 4 & 2.5987 & 2.6007 & $2\times 10^{-3}$\\
1 & 4 & 4 & 5.6515 & 5.6454 & $6\times 10^{-3}$\\
\end{tabular}
\end{ruledtabular}
\end{table}
\begin{figure}[p]
\includegraphics[scale=0.81]{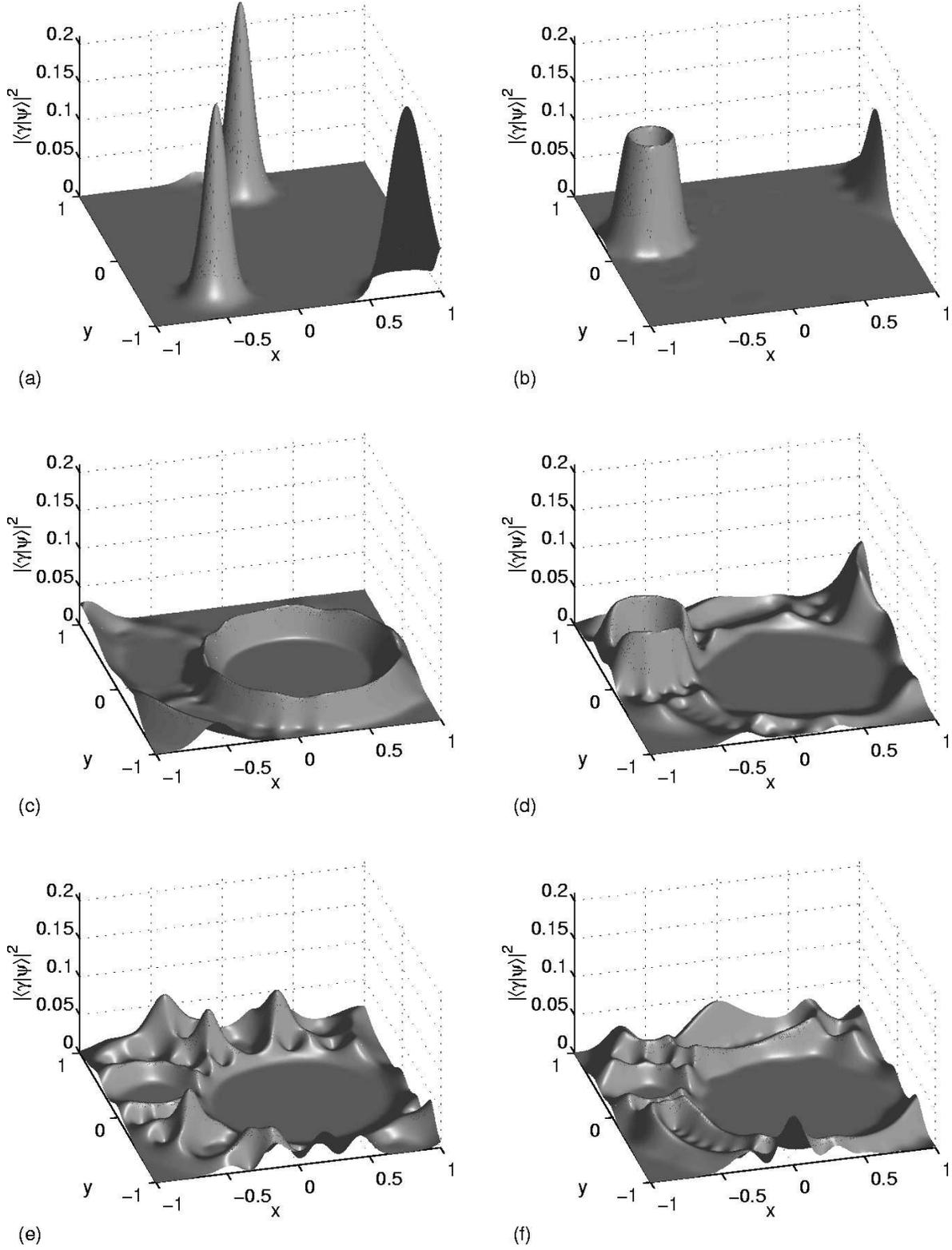}
\caption{The Husimi functions of selected eigenstates when 
$\mu=\omega=\pi(\sqrt{5}-1)$ and $j=100$. (a) ($m$,$n$,$p$)=(0,2,4), 
(b) (2,1,1), (c) (57,1,2), (d) (6,1,$1'$), (e) and (f) 
no association to a periodic orbit.}
\label{fig7}
\end{figure}
\begin{figure}[t]
\includegraphics[scale=0.92]{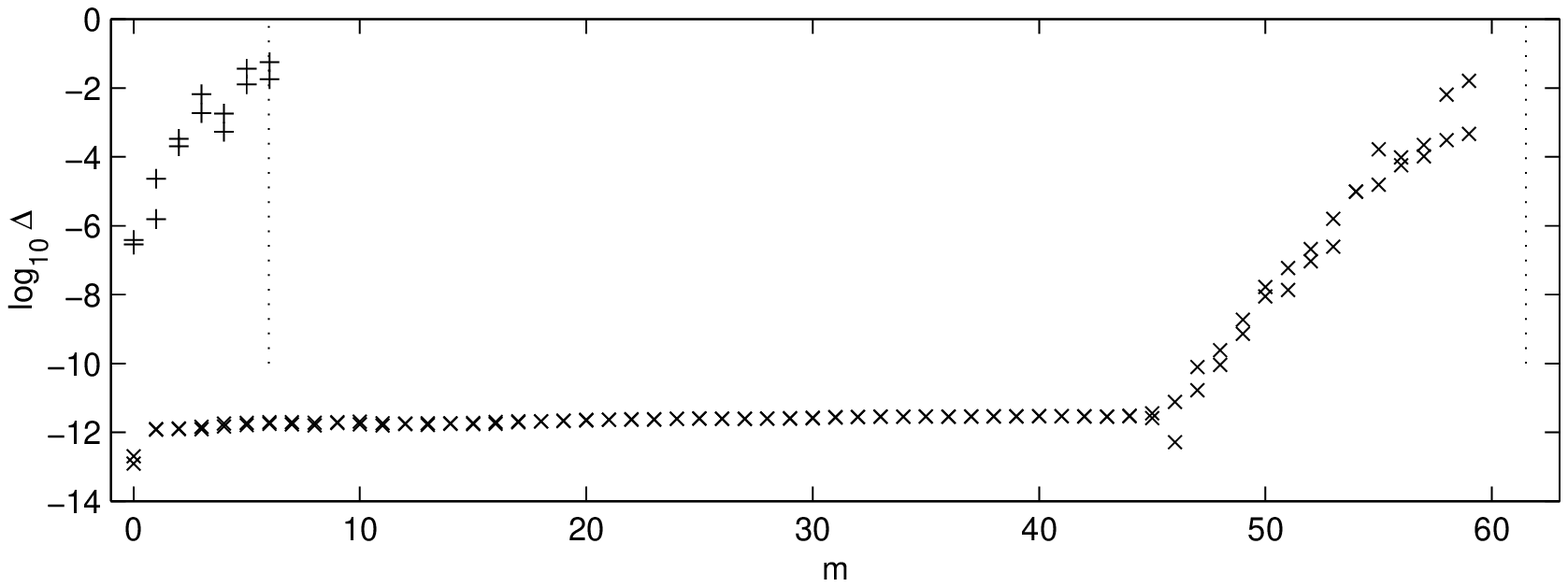}
\caption{The logarithm of the error in the semiclassical eigenphases 
versus $m$ for $p=1$ ($+$) and $p=2$ ($\times$).}
\label{fig8}
\end{figure}

Our semiclassical eigenphases (\ref{eigsc}), however, were found to provide good 
estimates of the quantum eigenphases. Some of the eigenphases for $j=100$ are 
tabulated in Table \ref{table2}. The quantum eigenphases $(\phi)_{qu}$ were 
calculated in the usual manner by diagonalizing the Floquet operator. The 
Husimi function $|\langle\gamma |\psi\rangle |^2$ of each eigenstate was then 
plotted in phase space to locate the periodic orbit from which it derives. 
Approximately 142 of the 201 quantum eigenphases could be classified as 
corresponding to a particular periodic orbit (120 for the period-2 orbit, 14 
for the period-1 pair, and 8 for the period-4 orbit). 

In Figure \ref{fig7} we have plotted the Husimi functions of selected 
eigenstates (refer to Figure \ref{fig3}(c) for the classical phase space). One 
of the four different ground states associated with the period-4 orbit is 
plotted in Figure \ref{fig7}(a). The Husimi functions of the three other ground 
states are very similar to this one. In the limit $j\rightarrow\infty$, each of 
these states will be in the form of a 4-fold `cat state'
\begin{equation}
\textstyle\frac{1}{2}e^{ik\pi/2}|\rho_1\rangle+\frac{1}{2}e^{il\pi/2}|\rho_2\rangle+\frac{1}{2}e^{im\pi/2}|\rho_3\rangle+\frac{1}{2}e^{in\pi/2}|\rho_4\rangle \label{cat}
\end{equation}
where $k$, $l$, $m$ and $n$ are integers, and $|\rho_k\rangle$ is a coherent 
state centered at the $k$-th iterate of the classical period-4 orbit. Hence, 
linear combinations of the four ground states with appropriate prefactors of 
the form $e^{ik\pi/2}/2$, will produce localized states on only 
one of the four points of the classical orbit. 

In Figure \ref{fig7}(b) we have plotted an eigenstate associated with the pair 
of period-1 orbits. This state is of higher order ($m=2$) and forms rings 
around the two classical fixed points. The period-1 orbits have degenerate 
actions (mod $2\pi$), and hence, their semiclassical eigenphases are equal. We 
have primed the second of each quantum number $p=1$ in Table \ref{table2} to 
emphasize this degeneracy. Note that the quantum eigenphases, however, are 
unequal. A small splitting allows `quantum tunnelling' \cite{dyrting} between 
the period-1 orbits. If we denote the two ground eigenstates by $|011\rangle$ 
and $|011'\rangle$, then the state $|011\rangle+|011'\rangle$ will be localized 
on one of the period-1 orbits, while the state $|011\rangle-|011'\rangle$ will 
be localized on the other. Only after $\pi/|(\phi_{011})_{qu}-(\phi_{011'})_{qu}|=3.05\times 10^7$ 
iterations of the quantum mapping, will each of these states have completely 
tunnelled their way across to each others period-1 point. 

In Figure \ref{fig7}(c) we have plotted an eigenstate associated with the 
period-2 orbit. This is another high-order eigenstate ($m=57$), and despite the 
large value of $m$, it is immediately recognizable as being derived 
from the period-2 orbit. An example of when our association becomes questionable
is given in Figure \ref{fig7}(d). We have classified this eigenstate as being 
derived from the period-1 pair. Figures \ref{fig7}(e) and (f) show examples of 
eigenstates which could not be classified as corresponding to a particular 
periodic orbit. Our semiclassical analysis affords no description of these 
eigenstates.

The logarithm of the error in the semiclassical eigenphases 
\[
\Delta\equiv |(\phi_{mnp})_{sc}-(\phi)_{qu}|
\]
versus $m$ is plotted in Figure \ref{fig8} for the period-1 ($+$) and period-2 
($\times$) orbits. This error decreases exponentially for decreasing $m$. The 
leveling out for $m\leq 46$ in the period-2 case is due to numerical error 
attributed to the use of double precision arithmetic. The dotted lines are our 
estimates for the maximum value of $m$ (\ref{M}) for which the 
semiclassical eigenphases will approximate a quantum eigenphase. In the period-1 
case $M=5.98$, whilst in the period-2 case $M=61.5$. In both cases our 
estimates give reasonable approximations.

\section{Conclusion}\label{sectionVIII}

As a first attempt at applying Gutzwiller's method of periodic orbit 
quantization to the mapping under consideration we have had some success. 
However we have also met great difficulties. We have found that our 
semiclassical approximation for the trace $\tr\hat{F}^n$ accumulates large 
errors upon increasing $n$. These errors decay for large $j$, however not at a 
rate rapid enough for resolution of the semiclassical spectrum. The errors were 
caused by an exponential proliferation of quantum effects, including ghosts, 
which were not accounted for in the simple semiclassical analysis. Despite this, 
our semiclassical eigenphase formula gave an accurate approximation to the 
quantum eigenphases associated with each periodic orbit. However, this accuracy 
is hardly surprising given the simplicity of our mapping in the neighborhoods of 
stable periodic orbits. The global dynamics is of more interest being of high 
complexity, similar to that of chaotic motion. The semiclassical analysis, 
however, could not describe eigenstates with support in these parts of the 
phase space. 

The accuracy of our semiclassical trace formula may be increased in three different ways: 
\begin{enumerate}
\item By including high-order terms in our approximation of the quantum matrix 
element (\ref{scmatrix}). These terms would need to be included if one were to 
study the semiclassical role of the unstable set. 
\item By including contributions from {\it all} saddle points of the integral 
(\ref{gint}). That is, include the exponentially small terms such as those from 
ghosts. 
\item By including high-order terms along with the first-order approximation 
(\ref{gauss}) for the integral (\ref{gint}). These terms would enhance the 
accuracy of our trace formula when the action $S$ becomes close 
to a multiple of $\pi$.
\end{enumerate}
However one must remember that, in general, an asymptotic series is 
expected to be divergent, and hence, including more terms may prove detrimental 
if $j$ is not further increased. 

The usefulness of Gutzwiller's method of periodic orbit quantization is 
debatable. There is no doubt that the theoretical insights offered are second 
to none. However the practicality of applying such a theory to accurately 
calculate the eigenphases, or even the traces themselves, is extremely limited. 
One must always keep in mind that the semiclassical analysis is only valid in 
the limit $j\rightarrow\infty$. How large $j$ needs to be, before the theory 
becomes an accurate one, will depend on the specific nature of each particular 
system. In our case, Gutzwiller's method has proven inadequate. This failure may be 
related to the particular dynamical nature of our mapping, which is of extreme 
ellipticity rather than the preferred hyperbolicity. However we believe that 
the problems encountered in this study will, in general, reappear in other 
studies whenever large powers of the Floquet operator are considered. The 
theory is too simplistic to describe the global complexity encountered in most 
chaotic systems. Only when $j$ is made large enough, such that some eigenstates 
will be dominated by the behavior in the immediate vicinity of a periodic 
orbit, will the theory produce accurate results. This view may be a depressing 
one, however it is hardly surprising given the particular nature of the 
Gutzwiller's approach. 

In conclusion, we are left with yet another, albeit cute, example of periodic 
orbit quantization in an ever expanding sea of literature on the subject. 
However we hope that our contribution is judged favorably, being qualitatively 
different from the much explored {\it hyperbolic} periodic orbit quantization.

\begin{acknowledgments}
The authors would like to thank Nico Temme for helping with the asymptotic 
anaylsis and also Catherine Holmes for further mathematical analysis.
\end{acknowledgments}

\appendix
\section{Asymptotics of ${\,}_2F_1\lowercase{(-j,1;j+1;z)}$}\label{appendixA}
The hypergeometric function under consideration is first replaced by another using a recurrence relation \footnote{Equation 2.8(33) on p.103 of \cite{erdelyi} with $a=-j$, $b=1$ and $c=j+1$.}
\begin{equation}
{\,}_2F_1\Big(-j,1;j+1;z\Big)=\frac{1}{2}+\frac{1-z}{2}{\,}_2F_1\Big(1-j,1;j+1;z\Big). \label{F} 
\end{equation}
This new hypergeometric function can be represented by the integral 
\begin{equation}
{\,}_2F_1\Big(1-j,1;j+1;z\Big)=j\int_0^1[(1-t)(1-zt)]^{j-1}dt. \label{I}
\end{equation}
When $|z|<1$ a complete asymptotic expansion can be found with the help of a quadratic transformation \footnote{Equation 2.11(34) on p.113 of \cite{erdelyi} with $a=1$, $b=1-j$ and $c=j+1$.}
\begin{eqnarray}
{\,}_2F_1\Big(1-j,1;j+1;z\Big) &=& \frac{1}{1+z}{\,}_2F_1\Bigg({\textstyle\frac{1}{2}},1;j+1;\frac{4z}{(1+z)^2}\Bigg) \label{a1}\\
&=& \frac{1}{1+z}\sum_{m=0}^{n-1}\frac{\Gamma(m+1/2)j!}{\sqrt{\pi}(j+m)!}\left(\frac{4z}{(1+z)^2}\right)^m+O(j^{-n}). \label{a2}
\end{eqnarray}
Taking the limit $n\rightarrow\infty$ the series only converges when $|4z/(1+z)^2|\leq 1$, however we still believe it gives a valid asymptotic expansion of (\ref{I}) if $|z|<1$. This is supported by considering a path of steepest descent in the integral. Letting
\[f(t)=(1-t)(1-zt)\]
we see that an asymptotic expansion of (\ref{I}) will involve a contribution from the endpoint $t=0$ where $f(t)=1$, and perhaps a contribution from the saddle point $t=1/2(1+1/z)$ where $f'(t)=0$. Letting $t=x+iy$, the path of steepest descent from $t=0$ is given implicitly by $\im(f(x+iy))=0$. When $|z|<1$ this path is
\[y(x)=\frac{-1}{2\im z}\left(1+\re z-2x-\sqrt{(1+\re z-2x)^2+4(1-|z|^2)x(1-x)}\right)\]
which connects $t=0$ to the other endpoint $t=1$ where $f(t)=0$. Along the contour $f(t)$ is monotonic, and hence, the expansion depends only on the local behaviour near $t=0$. On deriving the first terms in the expansion one finds agreement with (\ref{a2}). 

When $|z|>1$ the path of steepest descent from $t=0$ does not pass through $t=1$. It does, however, pass through the other zero at $t=1/z$. To investigate this case we start by rewriting (\ref{I}) as
\begin{eqnarray}
{\,}_2F_1\Big(1-j,1;j+1;z\Big) &=& j\int_0^{1/z}[(1-t)(1-zt)]^{j-1}dt+j\int_{1/z}^1[(1-t)(1-zt)]^{j-1}dt \nonumber\\
&=& \frac{j}{z}\int_0^1[(1-s)(1-s/z)]^{j-1}ds+j(-z)^{-j}(1-z)^{2j-1}\int_0^1[s(1-s)]^{j-1}ds \nonumber
\end{eqnarray}
where we have substituted $t=s/z$ in the first integral and $t=1/z+s(1-1/z)$ in the second ($s$ real). The contour in the second integral takes a path of steepest ascent from $t=1/z$ to the saddle point and then of steepest descent to $t=1$. This integral can be evaluated explicitly, giving an exact contribution from the saddle point. The contour in the first integral is not of steepest descent but now this integral takes the form of (\ref{I}). We have in fact rederived the linear transformation \footnote{Equations 2.9(2,6,10,25) on p.105 of \cite{erdelyi} with $a=2j$, $b=j$ and $c=j+1$.}
\begin{equation}
{\,}_2F_1\Big(1-j,1;j+1;z\Big)= \frac{1}{z}{\,}_2F_1\Big(1-j,1;j+1;1/z\Big)+2\frac{(j!)^2}{(2j)!}(-z)^{-j}(1-z)^{2j-1}. \label{a4}
\end{equation}
The asymptotic expansion of (\ref{I}) when $|z|>1$ can now be found by using (\ref{a2}). 

In the case $|z|=1$ ($z\neq-1$), the path of steepest descent is a straight line from $t=0$ to the saddle point and then to $t=1$. Hence we rewrite (\ref{I}) as
\begin{eqnarray}
{\,}_2F_1\Big(1-j,1;j+1;z\Big) &=& j\int_0^{\frac{1}{2}(1+\frac{1}{z})}[(1-t)(1-zt)]^{j-1}dt+j\int_{\frac{1}{2}(1+\frac{1}{z})}^1[(1-t)(1-zt)]^{j-1}dt \nonumber\\
&=& j(1+1/z)\int_0^{\frac{1}{2}}[1-4\cos^2(\theta/2)s(1-s)]^{j-1}ds+ \frac{(j!)^2}{(2j)!}(-z)^{-j}(1-z)^{2j-1} \nonumber
\end{eqnarray}
where $z=e^{i\theta}$ and we have substituted $t=s(1+1/z)$ in the first integral and evaluated the second. When $z\neq-1$ the first integral takes its maximum at the endpoint $s=0$. Again, the contribution from this point is (\ref{a2}). There is also a contribution from the endpoint $s=1/2$. This contribution is exponentially smaller, being $O(j^{-n})$ for all $n$, and will be ignored. If, however, $j$ is not large enough the resulting expansion will become inaccurate when $z$ is close to $-1$. Hence we take (\ref{a2}) with the result of second integral as our asymptotic expansion in this case.     

Summarizing the above results, when $z\neq-1$ the asymptotic expansion of our original hypergeometric function is
\begin{eqnarray}
{\,}_2F_1\Big(-j,1;j+1;z\Big)&=&\frac{1}{2}\Big[1-\sgn(1-|z|)\Big]\frac{(j!)^2}{(2j)!}\Bigg[\frac{(1-z)^2}{-z}\Bigg]^j+\frac{1}{2} \nonumber\\
&&\quad+\,\,\frac{1}{2}\frac{1-z}{1+z}\sum_{m=0}^{n-1}\frac{\Gamma(m+1/2)j!}{\sqrt{\pi}(j+m)!}\left(\frac{4z}{(1+z)^2}\right)^m+O(j^{-n})
\end{eqnarray}
and when $z=-1$ we have the exact result from (\ref{a4})
\begin{equation}
{\,}_2F_1\Big(-j,1;j+1;-1\Big)=\frac{(j!)^2}{(2j)!}2^{2j-1}+\frac{1}{2}.
\end{equation}
We have tested these asymptotics numerically and found good agreement. However
convergence is weak near $z=-1$.   

\section{A multiple integral}\label{appendixB}

Consider the relation
\[
I_n(a_1^{\phantom{*}},\dots,a_n^{\phantom{*}})\equiv\int\prod_{k=1}^n d^2z_k^{\phantom{*}}\,\,e^{a_k^{\phantom{*}}z_{k+1}^*z_k^{\phantom{*}}-z_k^*z_k^{\phantom{*}}}=\frac{\pi^n}{1-a_1^{\phantom{*}}\dots a_n^{\phantom{*}}}
\]
where we have assumed the integral is convergent, $z=x+iy$, $d^2z=dxdy$ and the 
subscript $n+1\equiv 1$. The proof is relatively simple and is achieved through 
induction. When $n=1$ the integral is trivial. We now assume the equality holds 
for $n=m$ and consider the case $n=m+1$ 
\begin{eqnarray*}
I_{m+1}(a_1^{\phantom{*}},\dots,a_{m+1}^{\phantom{*}}) & = & \int d^2z_1^{\phantom{*}}\dots d^2z_{m+1}^{\phantom{*}}\,\,e^{a_1^{\phantom{*}}z_2^*z_1^{\phantom{*}}+\dots+a_m^{\phantom{*}}z_{m+1}^*z_m^{\phantom{*}}+a_{m+1}^{\phantom{*}}z_1^*z_{m+1}^{\phantom{*}}-z_1^*z_1^{\phantom{*}}-\dots-z_{m+1}^*z_{m+1}^{\phantom{*}}} \\
& = & \pi \int d^2z_1^{\phantom{*}}\dots d^2z_m^{\phantom{*}}\,\,e^{a_1^{\phantom{*}}z_2^*z_1^{\phantom{*}}+\dots+a_{m-1}^{\phantom{*}}z_m^*z_{m-1}^{\phantom{*}}+a_m^{\phantom{*}}a_{m+1}^{\phantom{*}}z_1^*z_m^{\phantom{*}}-z_1^*z_1^{\phantom{*}}-\dots-z_{m}^*z_{m}^{\phantom{*}}} \\
& = & \pi I_m(a_1^{\phantom{*}},\dots,a_{m-1}^{\phantom{*}},a_m^{\phantom{*}}a_{m+1}^{\phantom{*}}) \\
& = & \frac{\pi^{m+1}}{1-a_1^{\phantom{*}}\dots a_m^{\phantom{*}}a_{m+1}^{\phantom{*}}}
\end{eqnarray*}
Hence the equality is true for all $n$. By making appropriate substitutions for 
$a_k$ and $z_k$ the multiple integral (\ref{gauss}) is solved.


\begin{thebibliography}{9}

\bibitem{gutzwiller}M.C.~Gutzwiller, J. Math. Phys. {\bf 8}, 1979 (1967); {\bf 10}, 1004 (1969); {\bf 11}, 1791 (1970); {\bf 12}, 343 (1971).

\bibitem{gutzwiller2}M.C.~Gutzwiller, {\it Chaos in Classical and Quantum Mechanics} (Springer-Verlag, New York, 1990).

\bibitem{EBK}A.~Einstein, Verhandl. Deut. Phys. Ges. {\bf 19}, 82 (1917); L.~Brillouin, J. Phys. Radium {\bf 7}, 353 (1926); J.B.~Keller, Ann. Phys. {\bf 4}, 180 (1958).

\bibitem{berry}M.V.~Berry and M.~Tabor, Proc. R. Soc. London A {\bf 349}, 101 (1976); J. Phys. A {\bf 10}, 371 (1977).

\bibitem{ozorio}A.M.~Ozorio~de~Almeida, {\it Hamiltonian Systems: Chaos and Quantization} (Cambridge University Press, Cambridge, 1988).

\bibitem{tomsovic}S.~Tomsovic, M.~Grinberg and D.~Ullmo, Phys. Rev. Lett. {\bf 75}, 4346 (1995); D.~Ullmo, M.~Grinberg and S.~Tomsovic, Phys. Rev. E {\bf 54}, 136 (1996).

\bibitem{sieber}M.~Sieber, J. Phys. A {\bf 30}, 4563 (1997).

\bibitem{schomerus}H.~Schomerus and F.~Haake, Phys. Rev. Lett. {\bf 79}, 1022 (1997).

\bibitem{main3}J.~Main and G.~Wunner, Phys. Rev. Lett. {\bf 82}, 3038 (1999).

\bibitem{haake}F.~Haake, {\it Quantum Signatures of Chaos} (Springer-Verlag, Berlin, 1991).

\bibitem{tabor}M.~Tabor, Physica D {\bf 6}, 195 (1983).

\bibitem{junker}G.~Junker and H.~Leschke, Physica D {\bf 56}, 135 (1992).

\bibitem{laksh3}A.~Lakshminarayan, Paramana J. Phys. {\bf 48}, 517 (1997).

\bibitem{keating}J.P.~Keating, Nonlinearity {\bf 4}, 309 (1991).

\bibitem{boasman}P.A.~Boasman and J.P.~Keating, Proc. R. Soc. London A {\bf 449}, 629 (1995).

\bibitem{laksh2}A.~Lakshminarayan, Phys. Lett. A {\bf 192}, 345 (1994).

\bibitem{sano}M.M.~Sano, J. Phys. A {\bf 29}, 6087 (1996).

\bibitem{eckhardt}B.~Eckhardt and F.~Haake, J. Phys. A {\bf 27}, 4449 (1994).

\bibitem{saraceno}M.~Saraceno and A.~Voros, Physica D {\bf 79}, 206 (1994).

\bibitem{laksh}A.~Lakshminarayan, Ann. Phys. {\bf 239}, 272 (1995).

\bibitem{toscano}F.~Toscano, R.O.~Vallejos and M.~Saraceno, Nonlinearity {\bf 10}, 965 (1997).

\bibitem{tanner}G.~Tanner, J. Phys. A {\bf 32}, 5071 (1999).

\bibitem{kus}M.~Ku\'s, F.~Haake and B.~Eckhardt, Z. Phys. B {\bf 92}, 221 (1993).

\bibitem{kus2}M.~Ku\'s, F.~Haake and D.~Delande, Phys. Rev. Lett. {\bf 71}, 2167 (1993).

\bibitem{scharf}R.~Scharf and B.~Sundaram, Phys. Rev. E {\bf 49}, R4767 (1994).

\bibitem{sundaram}B.~Sundaram and R.~Scharf, Physica D {\bf 83}, 257 (1995).

\bibitem{scharf2}R.~Scharf and B.~Sundaram, Phys. Rev. Lett. {\bf 77}, 263 (1996).

\bibitem{saito}K.~Saito and T. Nagao, J. Phys. Soc. Japan {\bf 68}, 1131 (1999).

\bibitem{bogomolny}E.B.~Bogomolny, Nonlinearity {\bf 5}, 805 (1992).

\bibitem{scott}A.J.~Scott, C.A.~Holmes and G.J.~Milburn, Physica D {\bf 155}, 34 (2001).

\bibitem{bohigas}O.~Bohigas, M.J.~Giannoni and C.~Schmit, Phys. Rev. Lett. {\bf 52}, 1 (1984).

\bibitem{perelomov}A.~Perelomov, {\it Generalized Coherent States and Their Applications} (Springer-Verlag, Berlin, 1986).

\bibitem{janszky}J.~Janszky, P.~Domokos and P.~Adam, Phys. Rev. A {\bf 48}, 2213 (1993).

\bibitem{macduffee}C.C.~MacDuffee, {\it Theory of Equations} (Wiley, New York, 1954).

\bibitem{main}J.~Main, V.A.~Mandelshtam and H.S.~Taylor, Nonlinearity {\bf 11}, 1015 (1998).

\bibitem{main2}J.~Main, Phys. Rep. {\bf 316}, 233 (1999).

\bibitem{wall}M.R.~Wall and D.~Neuhauser, J. Chem. Phys. {\bf 102}, 8011 (1995).

\bibitem{mandelshtam}V.A.~Mandelshtam and H.S.~Taylor, Phys. Rev. Lett. {\bf 78}, 3274 (1997); J. Chem. Phys. {\bf 107}, 6756 (1997).

\bibitem{dyrting}S.~Dyrting, G.J.~Milburn and C.A.~Holmes, Phys. Rev. E {\bf 48}, 969 (1993).

\bibitem{erdelyi}A. Erd\'{e}lyi et al., {\it Higher Transcendental Functions: Volume I} (McGraw-Hill, New York, 1953).


\end{thebibliography}
\end{document}